\documentclass[aps,prb,reprint,superscriptaddress]{revtex4-2}

\usepackage{graphicx}% 
\usepackage{caption}

\usepackage{dcolumn}% Align table columns on decimal point
\usepackage{bm}% bold math

\usepackage[mathlines]
{lineno}% Enable numbering of text and display math
\usepackage{svg}
\usepackage{amsmath}
\usepackage{subcaption}
\usepackage{ragged2e}
\usepackage{float}
\usepackage{natbib}
\usepackage[justification=justified,singlelinecheck=false]{caption}
\usepackage{hyperref}% add hypertext capabilities

\begin{document}
\preprint{APS/123-QED}

\title{Intrinsic limits of timekeeping precision in gene regulatory cascades}% Force line breaks with \\

\author{Juan Sebasti\'an Hern\'andez}
\affiliation{Department of Physics, Universidad de los Andes, Bogot\'a, Colombia}
\affiliation{Department of Electrical and Computer Engineering, University of Delaware, Newark, Delaware, USA}
\author{C\'esar Nieto}
\affiliation{Department of Electrical and Computer Engineering, University of Delaware, Newark, Delaware, USA}
\author{Juan Manuel Pedraza}
\affiliation{Department of Physics, Universidad de los Andes, Bogot\'a, Colombia}
\author{Abhyudai Singh}
\affiliation{Department of Electrical and Computer Engineering, Biomedical Engineering, Mathematical Sciences, Center of Bioinformatics and Computational Biology, University of Delaware, Newark, DE, USA}
\date{\today}% It is always \today, today,
             %  but any date may be explicitly specified

\begin{abstract}

Multiple cellular processes are triggered when the concentration of a regulatory protein reaches a critical threshold.~Previous analyses have characterized timing statistics for single-gene systems.~However, many biological timers are based on cascades of genes that activate each other sequentially.~Here, we develop an analytical framework to describe the timing precision of such cascades using a burst–dilution hybrid stochastic model.~We first revisit the single-gene case and recover the known result of an optimal activation threshold that minimizes first-passage-time (FPT) variability.~Extending this concept to two-gene cascades, we identify three distinct optimization regimes determined by the ratio of intrinsic noise levels and the protein dilution rate, defining when coupling improves or worsens timing precision compared to a single-gene strategy.~Generalizing to cascades of arbitrary gene length, we obtain a simple mathematical condition that determines when a new gene in the cascade can decrease the timing noise based on its intrinsic noise and protein dilution rate.~In the specific case of a cascade of identical genes, our analytical results predict suppression of FPT noise with increasing cascade length and the existence of a mean time that decreases relative timing fluctuations.~Together, these results define the intrinsic limits of timekeeping precision in gene regulatory cascades and provide a minimal analytical framework to explore timing control in biological systems.

% obtain a general criterion for evaluating the contribution of each gene to the cascade’s timekeeping performance.
% identify simple mathematical conditions that determine whether additional genes in a cascade improve or degrade the cascade's timekeeping performance.

% Old abstract
%Different cellular events are triggered when the product of a given gene reaches a critical threshold level.~Experimentally, many of these timing mechanisms couple multiple genes that activate one another sequentially, implementing a gene cascade.~In this article, we develop an analytical approach describing gene-cascade timing mechanisms.~We begin by analyzing a two-gene system and identify conditions under which gene coupling reduces timing noise compared to a single-gene setup.~We find that the optimal activation threshold that minimizes timing noise depends on the ratio of intrinsic noises of the involved genes, the target time, and the protein half-life.~We derive general expressions for a cascade of an arbitrary number of genes.~We establish conditions that each gene must meet to contribute to the reduction of noise throughout the cascade.~Finally, we estimate the lowest possible timing noise achievable in a cascade where all genes share identical stochastic properties.

\end{abstract}

\maketitle

%\tableofcontents

\section{Introduction}\label{intro}

Cells regulate physiological processes by activating and deactivating molecular mechanisms.~In many instances, activated molecules can activate other molecules through a regulatory cascade in which its final stage triggers a specific event.~Relevant examples include gene expression systems such as the synthesis of molecules involved in the development of the flagellar motor in \textit{E.~coli}~\cite{kalir2001ordering}, the lytic cycle of bacteriophages, viruses that infect bacteria, in which the components of new phages are synthesized sequentially~\cite{singh2014stochastic,mondal2024molecular}, and the sequential synthesis of cyclins to define transitions between different stages of the cell cycle~\cite{matthews2022cell}.~Similar mechanisms are also key in synthetic circuit design~\cite{english2021designing}.~Furthermore, other types of cascades with a similar structure include the sequential activation of molecules in signaling processes, such as mitogen-activated protein (MAP) kinases, in which MAPKKK molecules when reaching a threshold concentration activate MAPKK molecules that, in turn, activate MAPK molecules~\cite{avruch2007map}.~This signaling process is a common trigger in cellular events such as proliferation, differentiation, and apoptosis~\cite{ng2024role}.

In this article, we examine the timing process of gene expression cascades.~The process begins with the induction of the first gene, which produces a protein that activates a second gene once a certain threshold concentration (activation threshold) is reached.~This sequential cascade of multiple genes continues, with each gene activating the next, until the timekeeping process concludes when the last gene reaches its respective threshold, as depicted by Fig.~\ref{fig:cascade_cartoon}.~Since each step depends on the accumulation of molecules to a threshold, cascades inherently introduce a delay between signal initiation and the final response.~Therefore, with proper parameter settings, regulatory cascades can be used to manage the timing of molecular events~\cite{pedraza2007random,ali2022controlling}.~However, the stochastic nature of these molecular processes, such as the intrinsic random noise in the levels of proteins involved in cascades~\cite{eldar2010functional} and in their respective activation thresholds~\cite{co2017stochastic,biswas2016redundancy}, can disrupt the precision of these timers.~In fact, the malfunction of these mechanisms can dramatically affect cellular function~\cite{nachman2007dissecting}.~This motivates the study of how different architectures and variables within the cascade system can influence the accuracy of these molecular timers and which parameters can generate the most accurate timing~statistics~\cite{thattai2002attenuation}.

The performance of biomolecular timing systems is usually described in terms of the statistics of the first-passage time (FPT)~\cite{ghusinga2017first}, defined as the time interval between the initiation of a process and the achievement of the goal, typically reaching a threshold concentration.~To study the threshold crossing dynamics, multiple mathematical frameworks have been developed to study the stochasticity of the gene product levels~\cite{Elowitz_Levine_Siggia_Swain_2002,Thattai_vanOudenaarden_2001, Ozbudak_Thattai_Kurtser_Grossman_vanOudenaarden_2002, Pedraza_Paulsson_2008}.~Theoretical approaches include discrete frameworks such as the birth-death process and chemical master equation (CME) formulations~\cite{Rao_Waxman_Lin_Song_2025}, continuous-valued approaches based on stochastic differential equations~\cite{rijal2022exact}, and stochastic hybrid systems, with the burst–dilution representation being one of the simplest and easiest to manipulate analytically~\cite{singh2010stochastic}.~These models have been successfully applied to a wide range of regulatory systems, including self-regulation~\cite{ghusinga2017first, Cao_Qiu_Zhou_Zhang_2019, co2017stochastic, rezaee2023optimizing}, regulation by product degradation~\cite{rezaee2024controlling}, sRNA-mediated control~\cite{Ali_Prasad_Das_2025}, sequestration~\cite{ Biswas_Dey_Singh_2024} and phage lysis timing~\cite{kannoly2020optimum, kannoly2022optimal}, showing that these simplified representations produce useful, testable predictions despite biological approximations.~A fundamental result of these theories is that a shorter molecular half-life, related to gene product dilution or spontaneous degradation, reduces clock accuracy~\cite{ghusinga2017first,rezaee2023optimizing}.~As a result, in a context with non-zero degradation/dilution, there is an optimal  threshold that minimizes FPT variability~\cite{co2017stochastic,nieto2022threshold}.~This prediction has been experimentally verified in a single-gene cascade~\cite{kannoly2022optimal} and currently the existence of these optimal activation thresholds is not clear in cascades of an arbitrary number of genes.

To date, most FPT studies have derived exact analytical results to single-gene cascades~\cite{rijal2022exact}.~Other approximations estimated optimal parameters for self-regulation motifs and timers based on protein degradation~\cite{ghusinga2017first,nieto2022feedback,rezaee2024controlling,dey2021feedforward}.~Finally, there are numerical solutions to the two-gene system with self regulation~\cite{gupta2018temporal,gupta2020temporal}~Experimentally, multiple-gene cascade studies (notably the lytic pathway of $\lambda$-phage in \textit{E.~coli}) reveal that for a cascade of a given number of genes, as the cascade progresses, the FPT variance increases, while the coefficient of variation decreases~\cite{amir2007noise}.~This suggests that a higher temporal precision can be achieved with a large number of fast steps in the cascade.~However, it is still unclear what specific properties each cascade species (genes) must commit to enhance timing precision or how to estimate optimal activation threshold to minimize the FPT noise.~These open questions motivate the need for a general analytical description of timing in gene expression networks beyond one or two genes.

In this manuscript, by modeling the gene expression process as a burst-dilution hybrid stochastic system, we derive an analytical approximation for the noise in the cascade's FPT.~These formulas depend on the intrinsic noise and the activation threshold of each cascade gene.~Solving the optimization problem with a fixed mean FPT we obtain simple expressions for the optimal activation threshold levels, along with the conditions that a new gene must meet in order to improve the cascade timekeeping performance.~The article is structured as follows:~In Section~\ref{section2} we introduce the theoretical framework.~In Section~\ref{section3}, we revisit the optimization problem in a single-gene cascade.~In Section~\ref{section4}, we solve as an example the two-species cascade.~We explain how, depending on the intrinsic noise of gene expression and the protein dilution rate, a two-gene cascade could perform better than a single-gene cascade.~In Section~\ref{section5} we generalize our results to a cascade with an arbitrary number of species finding the ranges of intrinsic noise in which the gene decrease the FPT noise.~Finally, in Section~\ref{section6} we analyze cascades of identical genes finding the mean FPT that minimizes the FPT noise.~

\section{Model Formulation}\label{section2}

\begin{figure}[H]
    \centering  \includegraphics[width=\columnwidth]{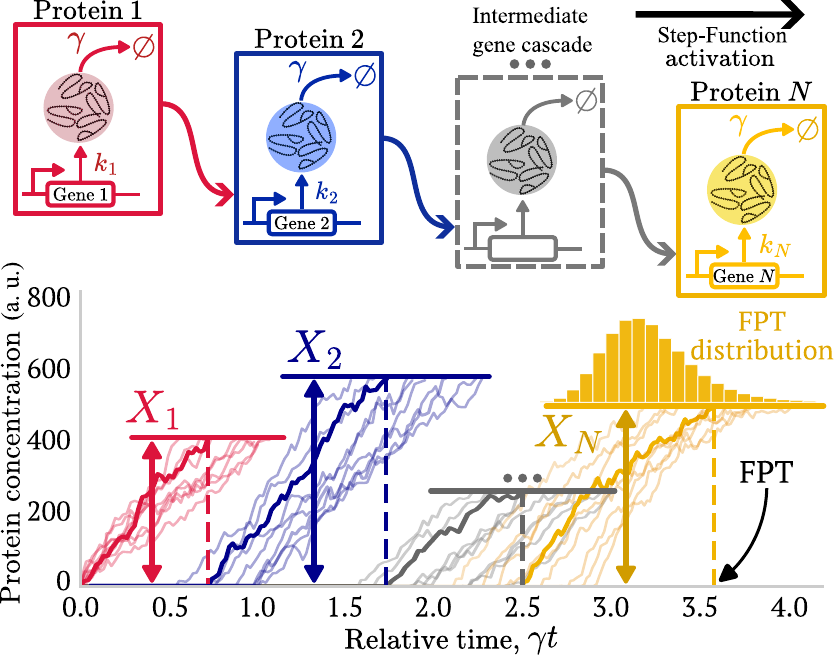}
    \caption{\small \textbf{Activation cascade of arbitrary length $N$ and resulting first-passage time (FPT) distribution}.~Schematic illustrating a sequential gene activation cascade (top) and the resulting stochastic dynamics of protein concentrations (bottom).~Each gene is activated by a \textit{Step-Function}: the expression of the activated gene starts just after the protein produced by the activating gene reaches its respective threshold ($X_1, X_2, \dots, X_{N-1}$).~Individual trajectories of protein concentrations illustrate how variability in timing arises from the inherent stochasticity of the system.~The histogram represents the distribution of first-passage times (FPT), measured when the final protein level surpasses the final threshold $X_N$, triggering the intracellular event.} % caption, usar esta convención
    \label{fig:cascade_cartoon} %etiqueta para llamar la gráfica desde el archivo
\end{figure}

The general timing system is depicted in Fig.~\ref{fig:cascade_cartoon}.~The process begins with all genes at basal level ($x_i=0)$ for simplicity.~The cascade begins when the first gene is activated ($t=0$).~Once its product $x_1$ (usually a protein) reaches the threshold level \( X_1 \), the second gene is activated, increasing the level of its product $x_2$.~This sequential activation continues: when the protein level of the preceding gene is above its corresponding threshold, the gene expression of the next gene in the cascade is activated.~The cascade continues in this way until the final protein concentration \( x_N \) surpasses the last threshold \( X_N \), triggering the intracellular event of interest.
 % Esta es la figura de este problema.
To mathematically describe gene expression, we introduce the \textit{burst-dilution}~\cite{kannoly2020optimum} model, which is formulated as a piecewise-deterministic Markov process.~Specifically, for the $i$-th gene we consider a continuously valued protein concentration, $x_i>0$, produced in random amounts called bursts.~This approximation is supported by several experiments showing that processes at every stage of protein production, including promoter activation, transcription, and translation, can be effectively approximated as bursting events in which a large number of proteins are produced in very short time intervals~\cite{Thattai_vanOudenaarden_2001, Ozbudak_Thattai_Kurtser_Grossman_vanOudenaarden_2002, Golding_Paulsson_Zawilski_Cox_2005, Cai_Friedman_Xie_2006, Yu_Xiao_Ren_Lao_Xie_2006, friedman2006linking, Pedraza_Paulsson_2008, shahrezaei2008analytical, Chong_Chen_Ge_Xie_2014, Fukaya_Lim_Levine_2016, Rodriguez_Larson_2020}.~Once the gene is active, protein bursts of the $i$-th transcription event arrive as a Poisson process with rate $k_i$.~Each burst increases $x_i$ by a random amount $b_i$, drawn from a general distribution with known first- and second-order moments $\langle b_i \rangle$ and $\langle b_i^2 \rangle$, respectively, where $\langle \cdot \rangle$ denotes the expected value.

In between bursts, we assume that $x_i$ decays primarily due to dilution through cell growth.~This assumption is valid for many bacterial proteins, which often exhibit half-lives of several hours~\cite{Gupta_Johnson_Cruz_Costa_Guest_Li_Hart_Nguyen_Stadlmeier_Bratton_2024}, making degradation negligible compared to dilution~\cite{singh2010stochastic, kannoly2020optimum, Zhang_Zabaikina_Nieto_Vahdat_Bokes_Singh_2025}.~This continuous decay can also be a good approximation to the spontaneous degradation if the amount of molecules is high enough~\cite{gillespie2009deterministic}.~In the case when dilution is the dominant source of decay, we can approximate that all proteins share the same dilution rate $\gamma>0$, which is equal to the cell's growth rate.~Therefore, gene products decay at an exponential rate following the differential equation ${\frac{dx_i}{dt}=-\gamma x_i}$.~Given the typical dilution rate, we define the dimensionless variable \textit{relative time} as $\gamma t$.~With this notation, during a relative time interval of $\gamma t = \ln(2)$, cell size doubles in size and protein level decreases by half.

%In between protein bursts, we assume that the gene product decays at an exponential rate following the differential equation \(\frac{dx_i}{dt}=-\gamma x_i\) with $\gamma$ known as the dilution rate.~Since spontaneous degradation is negligible for many bacterial proteins, whose half-lives often span several hours ~\cite{Gupta_Johnson_Cruz_Costa_Guest_Li_Hart_Nguyen_Stadlmeier_Bratton_2024}, protein level decay is mainly caused by dilution through cell growth~\cite{singh2010stochastic, kannoly2020optimum, Zhang_Zabaikina_Nieto_Vahdat_Bokes_Singh_2025}.~When dilution is the main source of protein level decay, we can assume that all proteins share the same dilution rate $\gamma$, which is equal to the growth rate.~With this approximation, we define the dimensionless variable $\gamma t$, denoted as relative time.~During a relative time interval $\gamma t = \ln(2)$, the cell size doubles and the protein level decays by half.

Having defined the production and dilution mechanisms for each gene product, we next describe how genes in the cascade interact.~In our model, the activation of gene $i$ by gene $i-1$ depends on the protein level $x_{i-1}$.~Although experimental activation is typically described by Hill functions of the activator protein level~\cite{Bintu_Buchler_Garcia_Gerland_Hwa_Kondev_Kuhlman_Phillips_2005, Garcia_Phillips_2011}, for mathematical tractability, we simplify this to a step-function activation.~In this approximation, the burst rate of gene $i$ remains zero until the gene product $x_{i-1}$ reaches its activation threshold $X_{i-1}$.~Upon reaching this threshold, the burst frequency instantly becomes $k_i$.

The quantity of interest in this work is the first-passage time (FPT), defined as the stochastic variable $T$ denoting the first instant at which the protein concentration of the last gene $x_N$ in the cascade reaches its corresponding threshold $X_N$ (see Fig.~\ref{fig:cascade_cartoon}).~Formally, the FPT is defined as follows:
{\small
\begin{align}
    T := \inf\{t \,:\, x_N(t) \geq X_N \mid x_i(0)=0,\ \forall i \in \{1,2,\ldots,N\}\}.
\end{align}}
In the case of a single-gene cascade, this definition reduces to the first time required for the protein concentration \(x_1\) to reach its threshold \(X_1\), without intermediate activation steps.~With all dynamical details specified, we can study the level of randomness of the stochastic variable $x_i$, which can be quantified by its statistical moments $\langle x_i^n \rangle$.~

\subsection*{Computing noise statistics}

Exact analytical expressions for the FPT distribution are generally difficult to obtain, even for simple biochemical networks~\cite{Rao_Waxman_Lin_Song_2025}.~To gain analytical tractability that would help us intuitively understand how the FPT depends on the main model parameters, we adopt an approximate approach based on moment dynamics formalism.~Specifically, we compute the low-order statistical moments of the protein concentrations and relate them to fluctuations in the FPT using a small-noise approximation~\cite{co2017stochastic}.~For the burst-dilution process, the moment dynamics of all gene products $x_i$ can be calculated using Dynkin's formula~\cite{singh2010stochastic} which describes the $n$-th moment of the $i$-th protein as a function of time is given by the following equation:

\begin{equation}\label{eq:moment_dynamics}
\frac{d\langle x_i^n \rangle}{dt} = \left\langle  A_i \left[(x_i + b_i)^n - x_i^n\right] \right\rangle 
- \left\langle \gamma x_i \cdot \frac{d}{dx_i}\left(x_i^n\right) \right\rangle,
\end{equation}

where step-function activation is implemented by defining the activation function \(A_i\) to take the value of the full activation rate \(k_i\) once the protein concentration of the activating gene, \(x_{i-1}\), reaches its corresponding threshold \(X_{i-1}\), and to be zero otherwise.~In the case of the first gene in the cascade, such activation is not required, and the gene is constitutively active from time \(t=0\), with \(A_1 = k_1\).

Using this formulation, we estimate the mean protein level \(\langle x_i \rangle\) and variance \(\sigma_i^2 := \langle x_i^2 \rangle - \langle x_i \rangle^2\), and define the squared coefficient of variation of the protein-level noise \(CV_i^2 = \sigma_i^2 / \langle x_i \rangle^2\) also known as noise.~This quantity is a dimensionless measure of relative noise used generally for positive-valued variables.~We relate $CV_N^2$ to the FPT noise, defined as \(CV_T^2 = \sigma_T^2 / \langle T \rangle^2\), via the equation~\cite{co2017stochastic}:
\begin{equation}\label{relating_protein_FPT}
CV_T^2 \approx \frac{\sigma_N^2}{\langle T \rangle ^2} \left( \frac{d \langle x_N \rangle}{dt} \right)^{-2} \bigg\rvert_{t = \langle T \rangle}.
\end{equation}

The approximation in Eq.~\eqref{relating_protein_FPT} is valid within a restricted regime.~It relies on a small-noise assumption, in which fluctuations of the protein concentration around its mean trajectory are sufficiently weak for linearization about the average dynamics to be justified.~Consequently, this relation provides accurate estimates when the threshold \(X_i\) lies below the steady-state protein level \(\bar{x}_i\)~\cite{co2017stochastic} and remains sufficiently separated from both zero and the steady state.~As illustrated in the inset of Fig.~\ref{fig:first}, deviations from stochastic simulations arise as the threshold approaches the steady state.~Despite these constraints, the small-noise approximation captures the dominant contribution to timing variability and has been applied across a broad range of biologically relevant network architectures~\cite{co2017stochastic,nieto2022threshold,rezaee2023optimizing, rezaee2024controlling}.

%\begin{subequations}\label{eq:moment_dynamics}
%\begin{align}
%\frac{d\langle x_i^n \rangle}{dt} &= \left\langle  A_i \left[(x_i + b_i)^n - x_i^n\right] \right\rangle 
%- \left\langle \gamma x_i \cdot \frac{d}{dx_i}\left(x_i^n\right) \right\rangle;\\[1ex]
%A_i &:= k_i\, {\displaystyle \Theta\left( {x_{i-1}- X_{i-1}} \right),\label{eq:activation_rule}} 
%\end{align}
%\end{subequations}

%for which an exponential distribution with $CV_T^2 = 1$ is typically used as a reference~\cite{ghusinga2017first}.~

% Vale la pena mencionar Dynkin's formula acá? En general, decir que así se obtienen los primeros momentos de todo y ya

%\section{Solution for unstable gene product ($\gamma\rightarrow0$)}

%In the appendix, we present the solution of \eqref{eq:moment_dynamics}

%\subsection{Simulation methods}

\section{Single-gene cascade}\label{section3}

%\subsection{Protein dynamics for single-gene cascade}

\begin{figure}[H]
\includegraphics[width=\columnwidth]{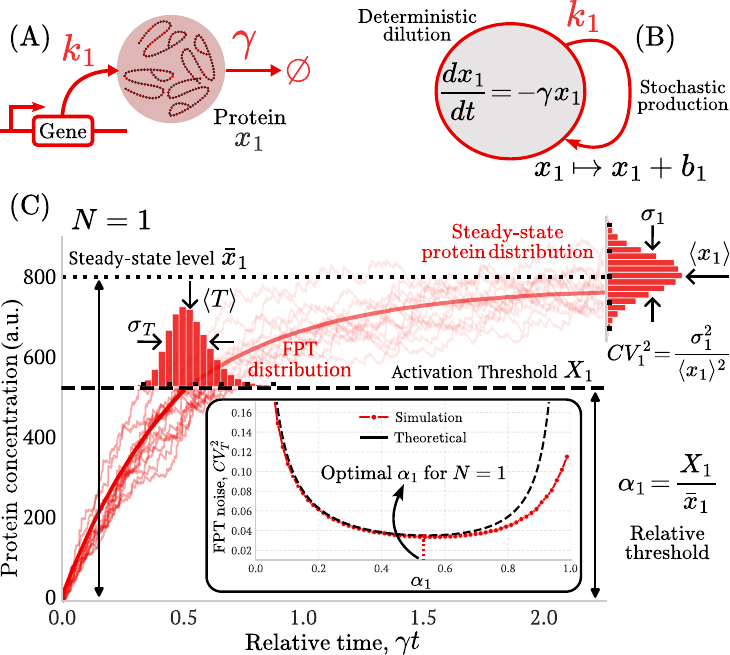}
    \caption{ \small \textbf{FPT statistics for a single-gene cascade.}~(A) Schematic of the burst-dilution model for a single gene that is constitutively expressed.~Proteins are produced in random bursts at rate $k_1$ and diluted continuously at exponential rate $\gamma$.~(B) Representation of gene expression as a stochastic hybrid process: discrete bursts events of size $b_1>0$ occur randomly in time, while protein levels decay exponentially between bursts.~(C) Example trajectories of protein accumulation from the initial state $x_1(0)=0$.~The event of interest is triggered when the protein level reaches the threshold $X_1$, which can be expressed relative to the protein steady-state $\bar{x}_1$ as $\alpha_1 = X_1 / \bar{x}_1$.~The first-passage time (FPT, $T$) is the random time required to reach this threshold, and its noise $CV_T^2$ depends on the steady-state protein noise $CV_1^2$, the relative threshold $\alpha_1$, and the dilution rate $\gamma$.~\textit{(Inset)} The $CV_T^2$ exhibits a U-shaped dependence on $\alpha_1$, with an optimal value at an intermediate relative threshold $\alpha_1^* \approx 0.55$~\cite{co2017stochastic, kannoly2020optimum}.~\textbf{Parameters:}~$\gamma = 0.05 \text{ min}^{-1}$, burst sizes taken from a geometric distribution with $\langle b_1 \rangle = 4$,  and $\bar{x}_1 = 800$, with $10^5$ replicates per point.} 
    \label{fig:first}  
\end{figure}

In this section, we will revisit the result of the existence of an optimal triggering threshold for a single-gene cascade.~Our framework is a simplification of more detailed approaches~\cite{co2017stochastic, kannoly2020optimum, ghusinga2017first, rijal2022exact}.~As discussed previously, the protein is produced in bursts with frequency $k_1$ and random size $b_1$ (Fig.~\ref{fig:first}A,~B).~Between consecutive bursts, the protein is diluted at an exponential rate $\gamma>0$.~In the case of a single-gene cascade, the FPT reduces to the time required for the protein concentration $x_1$ to reach its threshold $X_1$, without intermediate gene activation steps.~To solve the moments of the timing distribution, we start by estimating the protein dynamics of the system \eqref{eq:moment_dynamics} for $N=1$.~Fig.~\ref{fig:first}C shows an example of the typical protein dynamics for this single-gene model.~After solving the system \eqref{eq:moment_dynamics}, we find that the moments are given by:
\begin{subequations}
\begin{align}\label{mean_protein}
    \langle x_1 \rangle&=\bar{x}_1\left(1 - e^{-\gamma t}\right),\\
    \sigma^2_1 &= \frac{\langle b_1^2 \rangle}{2\langle b_1 \rangle}\bar{x}_{1}\left(1-e^{-2\gamma t}\right).\label{variance_single}
\end{align}
\end{subequations}

Here, $\bar{x}_{1}$ is the protein concentration  at steady state defined by the following expression:
\begin{equation}
    \bar{x}_{1} := \lim_{t \rightarrow \infty}{\langle x_1\rangle} = \frac{k_1\langle b_1 \rangle}{\gamma},
\end{equation}
which represents the typical protein level that cells reach in the absence of extrinsic factors.~This value will be the reference for the protein level, because its mean value satisfies $0\leq\langle x_1\rangle<\bar{x}_1$ in the small-noise approximation regime.~The definition of steady-state level is used to define the \textit{relative threshold} as:
\begin{equation}\label{eq:alpha}
    \alpha_1:= \frac{X_1}{\bar{x}_1},
\end{equation}
which measures how high $X_1$ is relative to $\bar{x}_1$ (as shown in Fig.~\ref{fig:first}C).~For simplicity, we will consider trigger thresholds $0<X_1<\bar{x}_1$, such that ${0<\alpha_1<1}$.~Finally, the protein concentration noise in the steady state is characterized by the squared coefficient of variation,

\begin{equation}\label{noise_protein}
    CV^2_1  := \lim_{t\rightarrow\infty}\frac{\sigma^2_1}{\langle x_1 \rangle^2} = \frac{\langle b_1^2 \rangle}{2\langle b_1 \rangle}\frac{1}{\bar{x}_1}.
\end{equation}

In this model, fluctuations in \(x_1\) arise solely from the stochasticity of bursty production, and therefore, $CV_1^2$ corresponds to the intrinsic gene expression noise.~For a fixed steady-state mean \(\bar{x}_1\), the intrinsic noise is determined by the burst statistics and does not depend explicitly on parameters such as \(k_1\) or \(\gamma\), which only enter through \(\bar{x}_1\).~

%Notice that for a fixed $\bar{x}_1$, $CV^2_1$ depends mainly on the burst statistics and not explicitly on other parameters such as $k_1$ or $\gamma$ which just define $\bar{x}_1$.

After solving the protein moments dynamics, we can simplify the FPT statistics, considering that we are in the small-noise regime.~We approximate the mean FPT $\langle T\rangle $ to the time $t$ it takes for the solution of $\langle x_1\rangle$ to reach $\langle x_1\rangle=X_1$ in the system \eqref{mean_protein}.~This solution is given by:

\begin{equation}\label{mean_fpt}
    \langle T \rangle \approx \frac{1}{\gamma} \ln{\left(\frac{1}{1 - \alpha_1}\right)},
\end{equation}
where \(\alpha_1\) is defined in Eq.~\eqref{eq:alpha} and denotes the relative threshold (see Fig.~\ref{fig:first}).~The mean first-passage time approximation in Eq.~\eqref{mean_fpt} breaks down for extreme values of \(\alpha_1\), overestimating the true mean as \(\alpha_1 \to 0\) and underestimating it as \(\alpha_1 \to 1\).~This deviation has been extensively analyzed in a recent study in a similar stochastic model (bursty birth-death)~\cite{grima_fpt}.~Nevertheless, despite this limitation, the approximation provides a tractable method for deriving analytical expression for the timing~statistics.

~Combining Eqs.~\eqref{relating_protein_FPT}, ~\eqref{variance_single}, \eqref{noise_protein}, and \eqref{mean_fpt}, we recover an expression for the FPT noise as a function of the relative threshold \( \alpha_1 \) (see~\cite{co2017stochastic, kannoly2020optimum} for a more detailed derivation):
\begin{align}\label{noise_fpt}
    CV_T^2 &\approx \frac{\sigma^2_1}{\langle T\rangle^2} \left( \frac{d \langle x_1 \rangle}{dt} \right)^{-2} \bigg\rvert_{t = \langle T \rangle}\nonumber\\
    &\approx CV_1^2 \cdot \frac{\alpha_1(2 - \alpha_1)}{(1 - \alpha_1)^2 \ln^2(1 - \alpha_1)}.
\end{align}

The noise described by Eq.~\eqref{noise_fpt} is a convex function of \( \alpha_1 \).~This implies the existence of an optimal threshold value that minimizes FPT noise.~We compare this dependence of $CV^ 2_T$ with the results of Monte Carlo simulations (Fig.~\ref{fig:first}C inset) showing that Eq.~\eqref{noise_fpt} overestimates the noise for values of \( \alpha_1 \) near 0 and 1.~Experimentally, the existence of this optimal threshold has been validated in the context of the lytic pathway of the $\lambda$-phage~\cite{kannoly2020optimum}.~In the following sections, we generalize this idea to activation cascades of arbitrary length.~First, we explore the two-gene cascade and identify the conditions under which optimal activation thresholds exist for both genes.

\begin{figure}[H]
    \centering \includegraphics[width=\columnwidth]{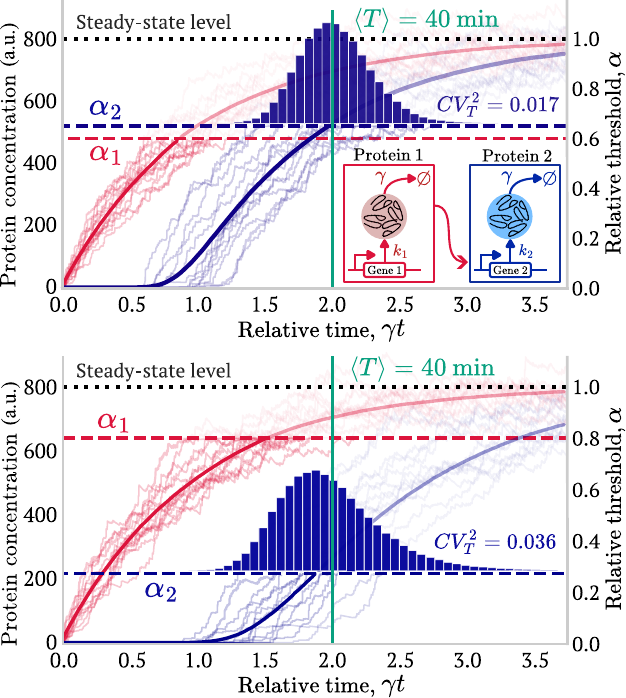}
    \caption{\small 
    \textbf{Threshold optimization in a two-gene cascade.}
    ~Solid lines are mean protein values, and semi-transparent lines are individual trajectories.~We explore how different relative threshold pairs $(\alpha_1, \alpha_2)$ affect the first-passage time (FPT) noise ($CV^2_T$) while maintaining a constant mean FPT ($\langle T\rangle$).~The optimization goal is to find the pair $(\alpha_1^*, \alpha_2^*)$ that minimizes $CV^2_T$ for a fixed $\langle T\rangle$.~\textbf{Parameters}:~$\gamma=0.05 \text{ min}^{-1}$, burst sizes taken from a geometric distribution with $\langle b_1\rangle=\langle b_2\rangle=4$, and $\bar{x}_1=\bar{x}_2=800$.~The target mean FPT is $\langle T\rangle=40 \text{ min}$ ($\gamma\langle T \rangle = 2$) (vertical line).~Histograms are from $10^5$ Monte Carlo replicates.
} % caption, usar esta convención
    \label{fig:cascada_dos} %etiqueta para llamar la gráfica desde el archivo
\end{figure}
\section{Two-gene cascade}\label{section4}

%\subsubsection{Solution for $\gamma\rightarrow 0$}

%A simple theoretical limit that can be studied is the solution for the case of $\gamma\rightarrow0$ (see Appendix \ref{no_dilution_section}).~This limit corresponds to a very slow dilution in relation to the timing scales.~As a main result, we show that in this case using multiple genes always increases the FPT noise, and the strategy that minimizes this noise is implementing the timekeeping process with a single species, choosing the gene with the least intrinsic noise.

We now consider the case of a two-gene cascade, in which gene 1 is activated at $t=0$.~The key assumption is that the second gene is activated with a \textit{step-function} mechanism (Fig.~\ref{fig:cascada_dos}) only after the protein product of the first gene $x_1$ reaches its threshold $X_1$.~This assumption is crucial because it allows us to model the entire process as a sequence of two independent events.~Therefore, the total time required for the final protein (from gene 2) to reach its threshold, $T$, can be expressed as the sum of two independent time intervals $T = T_1 + T_2$.~Here, $T_1$ is the time for the first gene product to reach its relative activation threshold $\alpha_1$.~$T_2$ is the subsequent time for gene product 2 to reach its relative threshold $\alpha_2$.~Using the small-noise approximation of Eq.~\eqref{mean_fpt}, the mean first-passage time (FPT) for the entire cascade is the sum of the mean FPTs for each gene, yielding the following approximation:
\begin{equation}\label{mean_two}
\langle T \rangle=\langle T_1 \rangle + \langle T_2 \rangle
\approx \frac{1}{\gamma} \ln{\left( \frac{1}{1 - \alpha_1} \cdot \frac{1}{1 - \alpha_2} \right)},
\end{equation}
where both genes are assumed to share the same dilution rate $\gamma$.~The optimization consists of minimizing the timing noise for a fixed mean FPT $\langle T\rangle$.~This constraint implies that the feasible range for each $\alpha_i$ is:

\begin{equation}\label{eq:range}    0\leq\alpha_i\leq1-e^{-\gamma\langle T\rangle}.
\end{equation}

Explicitly, for a given $\alpha_2$ and fixed $\langle T\rangle$, Eq.~\eqref{mean_two} can be used to obtain $\alpha_1$ in terms of $\alpha_2$:
\begin{equation}\label{eq:constraint}
    \alpha_1=1-\frac{e^{-\gamma\langle T\rangle}}{(1-\alpha_2)}.
\end{equation}

This relationship will be useful in the optimization problem.~Treating the cascade as a sequence of independent events also allows us to approximate the total FPT variance as the sum of the variances from each single-gene module, following Eq.~\eqref{relating_protein_FPT}.~Then, the resulting expression for the FPT noise in the two-gene cascade is approximated as:
\begin{subequations}
\label{noise_two_gene}  
\begin{align}
CV^2_T&\approx\frac{\sigma_{T_1}^2+\sigma_{T_2}^2}{(\langle T_1\rangle+\langle T_2\rangle)^ 2}\\&\approx\frac{CV^2_{1}\cdot\frac{\alpha_{1}\left(2 - \alpha_{1}\right)}{\left(1-\alpha_{1}\right)^2} + CV^2_{2}\cdot\frac{\alpha_{2}\left(2 - \alpha_{2}\right)}{\left(1-\alpha_{2}\right)^2}}{\ln^2{\left(\frac{1}{1-\alpha_{1}}\cdot\frac{1}{1-\alpha_{2}}\right)}}, 
\end{align}
\end{subequations}
where \( CV_{1}^2 \) and \( CV_{2}^2 \) correspond to the intrinsic noise of genes 1 and 2, respectively, as defined in \eqref{noise_protein}.~This expression can be solved after using Eqs. \eqref{mean_two} and \eqref{eq:constraint}.~Therefore, the FPT noise can be written in terms of the triggering threshold $\alpha_2$ as:
\begin{eqnarray}\label{noise_two}
CV_T^2 &\approx& \frac{1}{\left( \gamma \langle T \rangle \right)^2} \bigg[  CV_1^2 \cdot \left( (1 - \alpha_2)^2 e^{2\gamma \langle T \rangle} - 1 \right)\\
&&+ \:
CV_2^2 \cdot \frac{\alpha_2(2 - \alpha_2)}{(1 - \alpha_2)^2} 
\bigg];\quad 0\leq\alpha_2<1-e^{-{\gamma}\langle T\rangle}.\nonumber
\end{eqnarray}
The optimization reduces to determining the value of $\alpha_2$ that minimizes $CV_T^2$ with $\langle T \rangle$ fixed (Fig.~\ref{fig:cascada_dos}).~The nature of the optimal solution will depend on the intrinsic noise ratio $\frac{CV_2}{CV_1}$ as we will explain in the next section.

\subsection*{Conditions for improving the timekeeping precision (\texorpdfstring{$N=2$}{N=2})}

%the two-gene cascade timing noise function \eqref{noise_two} can become a monotonically increasing function of $\alpha_2$ (Fig.~\ref{fig:noise_two_gene_cascade}B, red region) and therefore is minimized in the lower limit $\alpha^*_2=0$.~Another option is that \eqref{noise_two} is a convex function of $\alpha_2$ (Fig.~\ref{fig:noise_two_gene_cascade}B, green region) and therefore has a nontrivial optimal $\alpha^*_2$.~The third option is that $CV_T^2$ is a monotonically decreasing function of $\alpha_2$  (Fig.~\ref{fig:noise_two_gene_cascade}B, blue region) and therefore it is minimized in the upper boundary $\alpha^*_2=1-e^{-{\gamma}\langle T\rangle}$.~These analytical results are further validated by Monte Carlo simulations, shown as circular markers in graph (B) of Figure \ref{fig:noise_two_gene_cascade}.
%The relative thresholds, $\alpha_1$ and $\alpha_2$, are by definition non-negative.~Furthermore, equation \eqref{mean_two} sets an upper limit on their values: $0 \leq \alpha_i < 1 - e^{-\gamma\langle T \rangle}$.~When one of the two thresholds reaches this upper bound ($1-e^{-\gamma\langle T \rangle}$) the entire time interval $\langle T \rangle$ is spent accumulating that specific protein to its threshold, and activation of the other gene occurs immediately.~If $\alpha^*_1$ or $\alpha^*_2$ predicted by \eqref{optimal_two_1} were to exceed this limit, the process would naturally take longer than $\langle T \rangle$, contradicting our initial assumption.~These conditions allow us to define three optimization regimes:

\begin{figure}[H]
    \centering \includegraphics[width=\columnwidth]{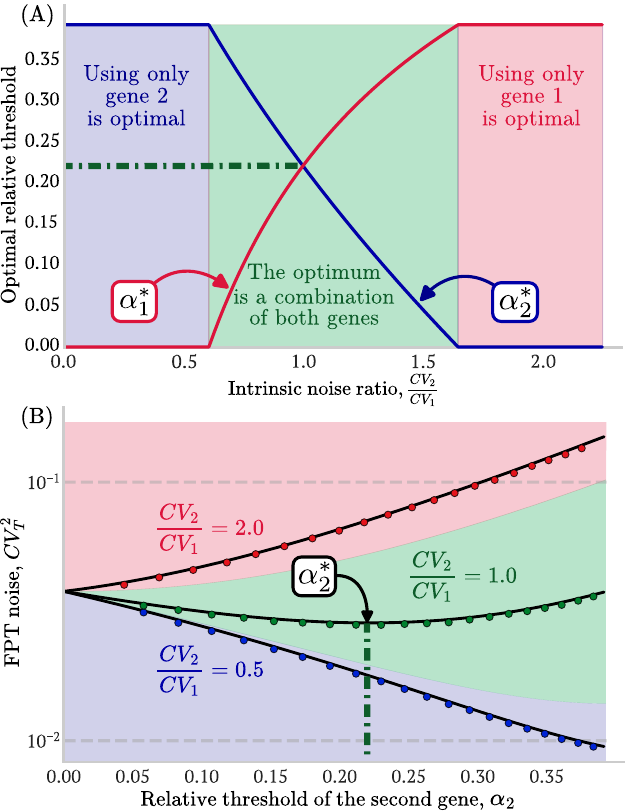}
    \caption{\small \textbf{Intrinsic noise sets a fundamental limit on timing precision.} (A) Optimal relative thresholds $\alpha_1^*$ (red) and $\alpha_2^*$ (blue) as functions of the intrinsic noise ratio $CV_2/CV_1$.~Minimizing the first-passage time (FPT) noise ($CV_T^2$) for a fixed mean FPT ($\langle T\rangle$) results in three optimization regimes: (i) using only gene 1 (red region), (ii) coupling both genes (green region), or (iii) using only gene 2 (blue region).~The horizontal green dashed line marks $CV_2/CV_1 = 1$, where both genes contribute equally.~(B) $CV_T^2$ as a function of $\alpha_2$ estimated for three representative cases, one per regime (black solid lines).~In the red (blue) regions, $CV_T^2$ increases (decreases) monotonically, yielding optimal threshold values at $\alpha_2^* = 0$ and $\alpha_2^* = 1 - e^{-\gamma\langle T\rangle}$, respectively.~In the intermediate green region, $CV_T^2$ shows a minimum within $0 < \alpha_2^* < 1 - e^{-\gamma\langle T\rangle}$, demonstrating that utilizing both genes reduces timing noise compared to a single-gene strategy.~This coupling region widens as the dilution rate $\gamma$ increases.~\textbf{Parameters}: $\gamma = 0.05~\text{min}^{-1}$,  $\langle T\rangle = 10~\text{min}$, $\bar{x}_1 = \bar{x}_2 = 1000$, burst sizes taken from a geometric distribution with~$\langle b_1\rangle = 4$.~Blue dots: ($\langle b_2\rangle$=0.625, $CV_2/CV_1 = 0.5$), green dots: ($\langle b_2\rangle$=4, $CV_2/CV_1 = 1.0$), and red dots: ($\langle b_2\rangle$=17.5, $CV_2/CV_1 = 2.0$).~Each data point consisted on $2\times10^5$ Monte Carlo replicates.}    \label{fig:noise_two_gene_cascade} 
\end{figure}

As shown in Fig.~\ref{fig:noise_two_gene_cascade}B, the timing noise function for a two-gene cascade \eqref{noise_two} can have three distinct behaviors within the allowed range of relative thresholds defined by Eq.~\eqref{eq:range}.~These regions depend on the value of the intrinsic noise ratio $\frac{CV_{2}}{CV_{1}}$, and determine whether coupling the two genes can improve timing precision.~These are:

\begin{itemize}
    \item \textbf{Using only gene 1 is optimal $\left(\frac{CV_2}{CV_1}>e^{\gamma\langle T\rangle}\right)$:} When the intrinsic noise of the second gene $CV^2_2$ is too high, $CV^2_T$ is a monotonically increasing function of $\alpha_2$ (see Fig.~\ref{fig:noise_two_gene_cascade} red region).~Therefore, the optimal solution lies in the boundary of the optimization region:
    \begin{subequations}\label{optimal_two_3}
\begin{eqnarray}
    \alpha_{1}^* &=& 1 - e^{-\gamma\langle T \rangle} \\ 
    \alpha_{2}^* &=& 0.
\end{eqnarray}
\end{subequations}
Thus, the best strategy is to use only the first gene for timing.~If the second gene is coupled in any way, it will increase timing noise.

    \item \textbf{The optimum is a combination of both genes $\left(e^{-\gamma\langle T\rangle} < \frac{CV_2}{CV_1} < e^{\gamma\langle T\rangle}\right)$:} In this scenario, $CV^2_T$ is a concave function of $\alpha_2$ (See Fig.~\ref{fig:noise_two_gene_cascade} green region).~The optimal solution consists of coupling both genes with the optimal activation thresholds:
\begin{subequations}\label{optimal_two_1}
\begin{eqnarray}
    \alpha_{1}^* &=& 1 - \left( \frac{CV_{1}}{CV_{2}} \cdot e^{-\gamma \langle T \rangle} \right)^{\frac{1}{2}} \\ 
    \alpha_{2}^* &=& 1 - \left( \frac{CV_{2}}{CV_{1}} \cdot e^{-\gamma \langle T \rangle} \right)^{\frac{1}{2}}.
\end{eqnarray}
\end{subequations}
Timing accuracy is improved by distributing the total mean FPT between both genes according to their intrinsic noise levels: the noisier the gene, the lower its relative activation threshold and therefore, the smaller its contribution to the accumulation time.

   \item \textbf{Using only gene 2 is optimal $\left( \frac{CV_2}{CV_1} < e^{-\gamma\langle T\rangle}\right)$:} When the first gene is too noisy, $CV_T^2$ decreases monotonically with $\alpha_2$ (Fig.~\ref{fig:noise_two_gene_cascade}, blue region).~The optimal solution again lies at the boundary:
        \begin{subequations}\label{optimal_two_2}
\begin{eqnarray}
    \alpha_{1}^* &=& 0 \\ 
    \alpha_{2}^* &=& 1 - e^{-\gamma\langle T \rangle}.
\end{eqnarray}
\end{subequations} Analogous to the red region, the best strategy is to use only the less noisy gene.~In this case, using only the second gene.
\end{itemize}

In Fig.~\ref{fig:noise_two_gene_cascade}A, we present the piecewise definition of the optimal thresholds combining expressions \eqref{optimal_two_3}, \eqref{optimal_two_1} and \eqref{optimal_two_2} and ~Fig.~\ref{fig:noise_two_gene_cascade}B~shows that the FPT noise approximation \eqref{noise_two} agrees well with the results from Monte Carlo simulations (circular markers).

%The three regimes defined by the ratio of the coefficients of variation ($\tfrac{CV_2}{CV_1}$) establish a natural boundary to determine when using a two-gene cascade is optimal to reduce FPT noise.~Although a gene cascade generally reduces FPT noise, if one gene is excessively noisy, the noisier gene contributes more variability to the total FPT than the coupling removes, making it optimal to exclude it.

%In particular, the condition \(e^{-\gamma\langle T\rangle} < \tfrac{CV_2}{CV_1} < e^{\gamma\langle T\rangle}\) defines the parameter region where the combination of both genes is optimal
The range of intrinsic noise ratios \(\tfrac{CV_2}{CV_1}\) for which gene coupling improves timing precision depends on the dilution rate \(\gamma\).~As the dilution rate increases, this interval (green region in Fig.~\ref{fig:noise_two_gene_cascade}) widens symmetrically in a logarithmic scale.~Consequently, higher dilution rates allow genes with increasingly different intrinsic noise levels to be combined effectively to reduce timing variability.~In contrast, in the no-dilution limit \((\gamma \rightarrow 0)\), the interval collapses to a single point, and coupling provides no benefit over using only the less noisy gene.~In this case, the optimal strategy reduces to selecting gene 2 if \(\tfrac{CV_2}{CV_1} < 1\), or gene 1 otherwise (see Appendix~\ref{no_dilution_section}).

\section{Cascade of an Arbitrary number  of genes}\label{section5}
We now extend these results to cascades with an arbitrary number of genes.~We consider a cascade of $N$ genes that are activated using the step-function approximation, as illustrated in Fig.~\ref{fig:cascade_cartoon}.~Applying the same logic as for the two-gene cascade, we treat this process as a sequence of $N$ independent events.~Then, the mean FPT and its noise can be written as:

\begin{subequations}
\begin{align}
\label{mean_general}
\langle T \rangle &\approx \frac{1}{\gamma}\ln{\left(\prod_{i=1}^{N}\frac{1}{1-\alpha_i}\right)}\\
CV^2_T &\approx\frac{1}{\left(\gamma\langle T\rangle\right)^2}\Bigg[\sum_{i=1}^{N-1}CV^2_{i}\cdot\frac{\alpha_{i}\left(2 - \alpha_{i}\right)}{\left(1-\alpha_{i}\right)^2} \nonumber\\&- \:CV^2_{N}\cdot \left(1-e^{2\gamma\langle T \rangle} \prod_{i=1}^{N-1}(1-\alpha_{i})^2\right)\Bigg].\label{noise_general}
\end{align}
\end{subequations}

Here, using the mean FPT constraint (Eq.~\eqref{mean_general}), we express the noise $CV^2_T$ as a function of the relative thresholds from $\alpha_{1}$ to $\alpha_{{N-1}}$.~Note that expressing the last relative threshold $\alpha_{N}$ in terms of the rest $N-1$ variables is arbitrary.~Due to the symmetry of the equations, choosing any $\alpha_{i}$ for this would be equivalent.

\subsection*{Minimization of FPT noise with a fixed mean}

The general \(N\)-gene analysis is analogous to the two-gene case: we determine the optimal relative thresholds by minimizing FPT noise under the fixed-mean $\langle T\rangle$ constraint.~After solving for the roots of the derivative of the expression  \eqref{noise_general}, we find that the optimal relative thresholds satisfy the following condition:
%todo: hacer el supplementary
\begin{figure*}[t]
   \centering \includegraphics[width=0.9\textwidth]{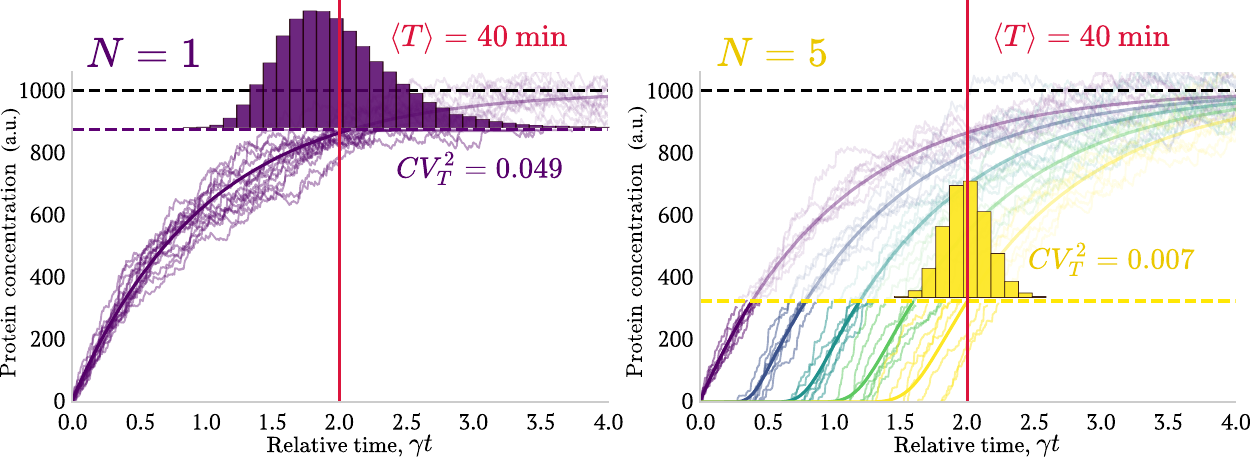}   \caption{\justifying \small \textbf{Longer cascades minimize noise for identical genes}.~Stochastic trajectories for cascades of $N$ identical genes with $N=1$ (left) and $N=5$ (right).~Thresholds were set to maintain a fixed FPT of $\langle T \rangle = 40~\text{min}$ (red vertical line).~Solid lines show mean protein values; semi-transparent lines show individual trajectories.~The FPT distribution for $N=5$ ($CV_T^2=0.007$) is substantially narrower than for $N=1$ ($CV_T^2=0.049$).~This demonstrates the strong noise reduction achieved by increasing $N$ with fixed $\langle T \rangle$, consistent with the theoretical $1/N$ scaling (Fig.~\ref{fig:tercera}).~\textbf{Parameters}: $\gamma=0.05 \text{ min}^{-1}$, $\bar{x_i}=1000$, burst sizes taken from a geometric distribution with $\langle b_i\rangle=4$.~Histograms were made with $1\times10^5$ replicates per simulation.
}
    \label{fig:ultima} 
\end{figure*} 
\begin{equation}\label{gradient}
    \frac{CV^2_{i}}{\left(1-\alpha^*_{i}\right)^3} = \left[CV^2_{N}e^{2\gamma\langle T \rangle}\prod_{j=1}^{N-1}\left(1-\alpha_{j}^*\right)\right]\prod_{k\neq i}^{N-1}\left( 1 -\alpha_{k}^*\right),
\end{equation}
with \( i, j, k \in \{1, 2, \dots, N-1\} \).

This expression yields \( N - 1 \) equations (one for every possible value of $i$) which, combined with \eqref{mean_general}, form a system of \( N \) independent equations.~Solving this system provides the \( N \) optimal relative thresholds $\alpha_i^*$ that minimize the noise in terms of the intrinsic noises $CV_i$.% Esto se podría quitar

~Dividing the \(i\)-th by the \(j\)-th equation results in the simple identity:
\begin{eqnarray}\label{recursion_relation}
\frac{CV_{i}}{1 - \alpha^*_{i}} = \frac{CV_{j}}{1 - \alpha^*_{j}}.
\end{eqnarray}

This relationship reveals a trend in optimal activation thresholds: genes with higher intrinsic noise should have their optimal relative threshold set closer to zero.~If intrinsic noise exceeds a critical level (dependent on the noise of the other genes), satisfying identity \eqref{recursion_relation} becomes impossible.~This implies that the noisy gene must be excluded from the cascade.
%In principle, Eg.~\eqref{recursion_relation} applies only for \( i, j \in \{1, 2, \dots, N-1\} \), since \(\alpha_N\) was eliminated to obtain Eq.~\eqref{noise_general}.~However, this choice is arbitrary.~If we had chosen any other $\alpha_{i}$ instead, the procedure and results would remain identical, the only difference being that the final expression would hold for all indexes except for the threshold that was eliminated.~This symmetry ensures that the equation derived in \eqref{recursion_relation} holds for all genes in the cascade.~Importantly, this recursion relation extends the intuitive result from the $N=2$ case: to achieve optimal timing precision, noisier genes are assigned smaller relative thresholds and therefore contribute less to the overall accumulation time.

%This freedom to choose any index introduces a symmetry, which consequently implies that the derived relation in \eqref{recursion_relation} must hold for all genes in the cascade.
%This symmetry (the freedom of choosing any index), therefore, implies that the derived relation \eqref{recursion_relation} must be valid for all genes in the cascade.

%The main concept we get from this expression is that the greater the intrinsic noise of the gene, the closer to one the ideal normalized threshold is.~

\subsection*{Conditions for improving the timekeeping precision ($N>2$)}

%The recursive relationship between the optimal thresholds \eqref{recursion_relation} depends explicitly on the intrinsic noise of each gene.~
Similarly to the two-gene case, given the fixed $\langle T\rangle$ constraint, there is a limited range of possible $\alpha_i$.~Using the recursive formula~\eqref{recursion_relation}, we obtain the analytical expression for the optimal thresholds as generalization of which is a generalization of Eq. \eqref{optimal_two_1}:
\begin{equation}\label{optimal_general}
    \alpha^*_{n} = 1-\left(e^{-\gamma\langle T \rangle}\cdot\prod\limits_{i=1}^{N}\frac{CV_n}{CV_{i}}\right)^{\frac{1}{N}},
\end{equation}
in which $CV_n$ corresponds to the intrinsic noise of the gene being evaluated, and with \( i \in \{1, 2, \dots, N\} \).~We also find that $\alpha^*_n$ lies within the achievable range \eqref{eq:range} as long as the intrinsic noise $CV_n$ of its corresponding gene falls within the interval:

%\begin{equation}\label{eq:multirange}
%   e^{-\gamma\langle T\rangle} <  \prod_{i=1}^{N}\frac{CV_i}{CV_n} < e^{(N-1)\gamma\langle T\rangle}
%\end{equation}

\begin{equation}\label{eq:multirange}
    e^{-\frac{(N-1)}{N}\gamma\langle T\rangle} < \frac{CV_n}{\left( \prod_{i=1}^{N}CV_i\right)^{\frac{1}{N}}} < e^{\frac{\gamma\langle T\rangle}{N}},
\end{equation}
which is the generalization to the range found for two-gene system. This result provides a general criterion for evaluating the contribution of each gene to the performance of the timekeeping of the cascade~Therefore, given any cascade with known intrinsic noise levels for its genes, condition \eqref{eq:multirange} offers a predictive tool to assess whether the inclusion of an additional gene could potentially improve or worsen the precision of the system.~Finally, it is also worth mentioning that the higher the dilution rate, the wider the range of tolerable noise for a gene to be useful in the cascade.
%Stochastic trajectories for cascades of identical genes with $N=1$ (left) and $N=5$ (right) are shown.~In both cases, thresholds were chosen such that the mean first-passage time is fixed at $\langle T \rangle = 40~\text{min}$ (red vertical line).~Solid lines show mean protein values, while semi-transparent lines represent individual stochastic trajectories.~Histograms display the resulting FPT distributions.~For $N=1$, the distribution is broad and skewed ($CV_T^2=0.049$), while for $N=5$ the distribution narrows substantially ($CV_T^2=0.007$).~This comparison illustrates the strong noise reduction obtained by increasing the number of identical genes while keeping the mean FPT constant, in agreement with the theoretical $1/N$ scaling shown in Fig.~\ref{fig:tercera}.~Simulations were performed with $\gamma=0.05 \text{ min}^{-1}$, steady-state concentration $\bar{x}=1000$, and a mean burst size of 4.~Each simulation included $1\times10^5$ replicates.

%, providing a fundamental guideline for optimizing the cascade’s design.

\section{Identical gene cascades}\label{section6}

For a cascade of $N$ identical genes and  fixed~$\langle T\rangle$, each with the same intrinsic noise \(CV_1^2\), the optimal solution has all gene products sharing the same activation threshold \(\alpha^* = 1 - e^{-\gamma \langle T \rangle / N}\).~Substituting this expression into Eq.~\eqref{noise_general}, we obtain a compact result for the FPT noise at the optimal threshold:
\begin{equation}\label{noise_identico}
    {CV_T}^2_{(\alpha^*)} \approx CV_1^2N\frac{e^{2\frac{\gamma\langle T\rangle}{N}}-1}{(\gamma\langle T\rangle)^2}.
\end{equation}

The expression \eqref{noise_identico} shows how the FPT noise decreases as the number of genes ($N$) increases.~To visualize this noise suppression effect, we present Fig.~\ref{fig:ultima} showing representative stochastic gene expression trajectories and their resulting FPT distributions for two different values of $N$, while maintaining $\langle T\rangle$ fixed.

These stochastic trajectories provide an intuitive understanding of how the timing variability shrinks.~In Fig.~\ref{fig:tercera} we provide a quantitative comparison.~Here, the analytical results predicted by Eq.~\eqref{noise_identico} are compared with stochastic simulations by plotting the FPT noise as a function of the relative mean FPT $\gamma \langle T \rangle$ for various values of $N$.~This comparison validates our analytical prediction \eqref{noise_identico} across different cascade lengths and dilution regimes.
 %The expression \eqref{noise_identico} illustrates how the FPT noise decreases as the number of genes increases.~To visulize this effect, we plot Fig.~\ref{fig:ultima} showing stochastic gene expression trajectories and their resulting FPT distribution for two different values of N and the same mean FPT.~Furthermore, the analytical results predicted by Eq.~\eqref{noise_identico} are compared with simulations in Figure \ref{fig:tercera}, showing the FPT noise as a function of $\gamma \langle T \rangle $ for different values of $N$.

%Expression \eqref{noise_identico} can also used to estimate fundamental limits to noise suppression for a given  $\langle T\rangle$.~Using the property that the FPT noise is a decreasing function of $N$, we can estimate the limit of very large number of identical genes, obtaining minimum achievable FPT noise, ideally, for a given mean timing $\langle T\rangle$:
%\begin{equation}\label{eq:minnoise}
%\lim_{N\rightarrow\infty}{CV_T}^2_{(\alpha^*)}=\frac{\langle b_1^2\rangle}{\langle b_1\rangle^2}\frac{1}{k_1\langle T\rangle}.
%\end{equation}
%Notice that this minimum noise limit is independent of the dilution rate $\gamma$.~It precisely corresponds to the limit where the dilution rate approaches zero ($\gamma \rightarrow 0$) (see Appendix \ref{no_dilution_section}).~In this limit, the optimal shared threshold ($\alpha^*$) is very low, which means that proteins do not dilute significantly before reaching their respective activation thresholds.
\subsection*{An optimal $\langle T\rangle$ minimizes the FPT noise given $N$}

\begin{figure}[H]
    \centering \includegraphics[width=\columnwidth]{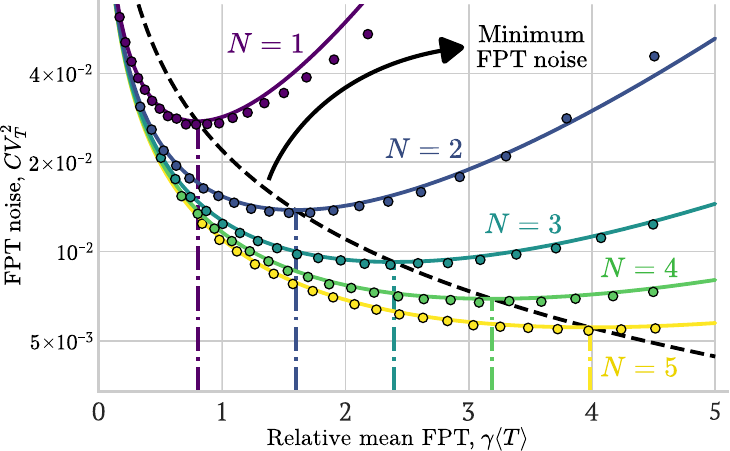}
    \caption{\small 
    \textbf{Optimal noise reduction in identical gene cascades for fixed $\gamma$.} $CV_T^2$ as a function of $\gamma\langle T\rangle$ for cascades composed of $N=1$ to $5$ identical genes.~Theoretical approximations using Eq.~\eqref{noise_identico} (solid lines) are compared with simulation results (circular markers).~Each curve shows a minimum noise level at an optimal relative time, $\gamma\langle T\rangle^*$, which is indicated by the vertical lines (color-coded by $N$).~The minimum FPT noise (dashed black line) is given by \eqref{ideal_identicos}.~As $N$ increases, both the overall noise level and its minimum decrease, scaling approximately as $1/N$.~This demonstrates that cascades of identical genes effectively suppress timing variability.~\textbf{Parameters}: $\gamma = 0.05 \text{ min}^{-1}$, $\bar{x_i} = 1000$, burst sizes taken from a geometric distribution with $\langle b_i\rangle=4$.~All genes shared the same relative threshold for each simulation.~Each data point used $2\times10^5$ Monte Carlo replicates.
}
    \label{fig:tercera}
\end{figure}

By relaxing the constraint over a fixed $\langle T\rangle$, it is possible to observe that, given $N$, the expression \eqref{noise_identico} is a convex function of $\langle T\rangle$ (see Fig.~\ref{fig:tercera}).~This noise reaches a minimum value at an optimal $\gamma\langle T\rangle^*$ that can be obtained by finding the roots of the derivative of the expression \eqref{noise_identico}.~This optimization results in the optimal  $\gamma \langle T\rangle^*$ which satisfies the equation:

\begin{equation}\label{tau_minimo}
    \left(1-\frac{\gamma\langle T\rangle^*}{N}\right)e^{\textstyle \frac{2\gamma\langle T\rangle^*}{N}}=1.
\end{equation}

This is a transcendental equation with no closed-form solution.~However, it can be shown that the solution has the form $\gamma \langle T\rangle^* = N\cdot\gamma \langle T\rangle^*_1$, where $\langle T\rangle^*_1$ is the solution of Eq.~\eqref{tau_minimo} when $N=1$, and is approximately equal to $\gamma\langle T\rangle^*_1\approx0.797$.~This optimal value $\langle T \rangle^*_1$ has been experimentally observed in the $\lambda$-phage lysis system as shown in~\cite{kannoly2020optimum}.

In general, the optimal $\gamma \langle T\rangle^*$ is a linearly increasing function of $N$ as can be visualized in Fig.~\ref{fig:tercera} which shows how consecutive optimal values of $\langle T\rangle^*$ have equal spacing between them (vertical color-coded lines). This result allows us to derive an expression for the minimum achievable FPT noise with unlimited $\langle T\rangle$ and fixed $\gamma$ as a function of $N$:

\begin{align}
\label{ideal_identicos}            \left[{CV_T}^2_{(\alpha^*)}\right]_{\min} &\approx \frac{CV^2_{1}}{\left(1-\gamma\langle T\rangle^*_1\right)\gamma\langle T\rangle^*_1}\frac{1}{N}\nonumber\\
&\approx 6.181\frac{CV^2_{1}}{N},
\end{align}
which is represented in Fig.~\ref{fig:tercera} by a dashed black line.~The minimum values of the curves of $CV^2_T$ vs $\langle T\rangle$ agree well with both simulations and the minimum predicted by the analytical expression~\eqref{ideal_identicos}.~

\section{Discussion}\label{section7}

In this work, we develop an analytical framework for analyzing the timing precision of regulatory cascades, extending the previous first-passage time (FPT) analysis from single-gene systems to cascades of arbitrary length.~For a single-gene cascade (Section \ref{section3}), we revisit the known result that for a given $\gamma>0$, an optimal relative threshold minimizes the FPT noise.~Additionally, this minimal noise also leads to an optimal value of $\gamma\langle T\rangle$, which has been experimentally observed previously~\cite{kannoly2020optimum}.~For a two-gene cascade (Section \ref{section4}), we show how $\gamma$ defines the range of possible intrinsic noise values of the second gene for which coupling it reduces timing noise relative to using only the least noisy gene.~A greater $\gamma$ increases this range, allowing noisier genes to actively contribute to timekeeping performance.~In a general cascade with more than two genes (Section \ref{section5}), a similar allowable range of intrinsic noises is observed (Eq. \eqref{eq:multirange}).~A new gene in the cascade will reduce timing noise if its intrinsic noise, relative to the geometric mean of the other noises, lies within a range dependent on $\gamma$.~Finally, we solve the case of a cascade of $N$ identical genes, finding that this system shares properties with the single-gene cascade (Section \ref{section6}).~For example, there is both an optimal threshold and an optimal $\langle T\rangle$, each dependent on $\gamma$, that minimize the timing noise for a given value of $N$.~In this identical-gene case, our analytical solution indicates that timing noise can be suppressed by increasing the number of cascade stages (Eq. \eqref{ideal_identicos}).

The central strength of our approach is its analytic tractability: we obtain closed-form expressions for mean timing, noise levels, and optimal thresholds that can be generalized systematically across cascade architectures.~With this advantage, we identify simple mathematical conditions (such as inequalities involving intrinsic noise ratios) that determine whether additional genes in a cascade improve or degrade timing precision.~Beyond providing intuitive insights into the role of cascade parameters, these results offer baseline predictions against which more complex models can be compared.

At the same time, our framework necessarily relies on simplifying assumptions that would limit its direct applicability.~First, much of our analysis is based on the small-noise approximation, which ensures that mean trajectories are well defined and that variability can be treated as a perturbation.~This approximation is convenient and captures the essential dependence of noise on thresholds and burst parameters, but it may break down in regimes of strong stochasticity, particularly for low copy-number proteins (for instance, if the threshold level is very low) where rare events dominate and deterministic estimates of mean first-passage times have been shown to lose accuracy~\cite{grima_fpt}.

Second, and perhaps more critically, our analysis focuses exclusively on intrinsic noise arising from stochastic expression and dilution of individual gene products.~While this restriction allows us to derive general and transparent expressions, it omits extrinsic fluctuations (such as variability in cell growth rates~\cite{Klumpp_Zhang_Hwa_2009, Keren_vanDijk_Weingarten-Gabbay_Davidi_Jona_Weinberger_Milo_Segal_2015, Weiße_Oyarzún_Danos_Swain_2015}, ribosome or global resource availability~\cite{pavlou2025single}, and cell-cycle stage~\cite{Sukys_Grima_2025}) that are often the dominant contributors to timing variability in single-cell experiments.~Such extrinsic noise would correlate fluctuations across cascade stages and could impose a lower bound on achievable precision, regardless of the optimal threshold placement suggested by intrinsic-noise arguments.

Therefore, our results should be interpreted as defining the intrinsic limit of timing precision in cascades: the minimal variability achievable if extrinsic noise were absent or negligible.~This perspective is valuable because it clarifies the specific contribution of intrinsic fluctuations and highlights the architectural constraints that govern them.~Future work should extend this analytic framework to include correlated extrinsic noise sources and graded activation functions, which would bring the theory closer to biological reality and allow more direct comparisons with experimental measurements.
\section*{Acknowledgements}

This work is supported by NIH-NIGMS via grant R35GM148351.

\appendix
\section{Solution for the no dilution case ($\gamma\rightarrow0$)}\label{no_dilution_section}

In this section, we consider the limit of the system given by \eqref{eq:moment_dynamics} when ${\gamma\rightarrow0}$, which corresponds to the case in which the gene product does not get degraded or diluted.~Biologically, this limit is relevant in very fast cascades when $\langle T\rangle\ll1/\gamma$.~The equations describing the moment dynamics are derived from the system \eqref{eq:moment_dynamics} and can be expressed as follows:
\begin{subequations}\label{eq:moment_dynamics2}
\begin{align}
\frac{d\langle x_i^n \rangle}{dt} &= \left\langle  A_i \left[(x_i + b_i)^n - x_i^n\right] \right\rangle ;\\[1ex]
A_i &:= k_i\, {\displaystyle \Theta\left( {x_{i-1}- X_{i-1}} \right).~\label{eq:activation_rule2}} 
\end{align}
\end{subequations}

Here, we consider that each gene activates the other through a step function as explained in the main text.~Similarly, we can minimize $CV^2_T$ for a fixed $\langle T\rangle$.

\subsection*{Single-gene cascade for $\gamma\rightarrow0$}
~In this system, a gene is constitutively expressed and the event of interest is triggered when a threshold concentration $X_1$ is reached.~A single gene with burst rate $k_1$ and random burst size with first- and second-order moments $\langle b_1\rangle$ and $\langle b^2_1\rangle$, respectively, has moments obtained by solving \eqref{eq:moment_dynamics2}:
\begin{subequations}
\begin{align}\label{eq:momsol}
 \langle x_1\rangle&=k_1\langle b_1\rangle t\\
  \sigma^2_1&=k_1\langle b^2_1\rangle t,\\
   CV^2_1&=\frac{\langle b^2_1\rangle}{k_1\langle b_1\rangle^2t}\label{eq3c}
\end{align}
\end{subequations}
in which the time $t$ is measured from gene activation.~The FPT noise can be related to intrinsic protein noise via the equation~\eqref{relating_protein_FPT}.~Using this approximation, we obtain the timing moments:
\begin{subequations}\label{eq:momsol2}
\begin{align}
\langle T_1\rangle&\approx\frac{X_1}{k_1\langle b_1\rangle}\\
\sigma^2_{T_1}&\approx\frac{\langle b^2_1\rangle}{k_1^2\langle b_1\rangle^3}X_1 =\frac{\langle b^2_1\rangle}{\langle b_1\rangle^2}\frac{\langle T_1\rangle}{k_1}.
\end{align}
\end{subequations}
Notice that for a fixed mean timing $\langle T_1\rangle$ and the gene expression parameters $k_1$,$\langle b_1\rangle$ and $\langle b^2_1\rangle$, the timing noise $CV^ 2_{T}=\sigma^2_{T_1}/\langle T_1\rangle^2$ is fixed and therefore, there is not optimization problem.

\subsection*{Two-gene cascade for $\gamma\rightarrow0$}

The two-gene cascade is composed of two timing events.~The first is the activation of the second gene once the first gene reaches its threshold $X_1$ and the event triggering when the product of the second gene reaches its threshold $X_2$.~Using the step-function activation, both events are considered independent and consecutive.~Therefore, the first gene has mean timing $\langle T_1\rangle$ while the second has mean timing $\langle T_2\rangle$ such that:
\begin{subequations}\label{eq:app4}
    \begin{align}
        \langle T\rangle&=\langle T_1\rangle+\langle T_2\rangle\\
        &\approx\frac{X_1}{k_1\langle b_1\rangle}+\frac{X_2}{k_2\langle b_2\rangle}.
    \end{align}
\end{subequations}
This constraint couples the activation thresholds $X_1$ and $X_2$ through the equation:
\begin{equation}\label{eq:x2}
    X_2=k_2\langle b_2\rangle\left(\langle T\rangle-\frac{X_1}{k_1\langle b_1\rangle}\right),
\end{equation}
with $\frac{X_1}{k_1\langle b_1\rangle}\leq\langle T\rangle$.~Using the system \eqref{eq:momsol2}, the timing variance $\sigma^2_T$ for the combined system is given by:
\begin{subequations}\label{eq:app6}
\begin{align}
\sigma^2_T&=\sigma^2_{T_1}+\sigma^2_{T_2}\\
         &\approx\frac{\langle b^2_1\rangle}{k_1^2\langle b_1\rangle^3}X_1+\frac{\langle b^2_2\rangle}{k_2^2\langle b_2\rangle^3}X_2,   
\end{align}
\end{subequations}

which, using the constraint \eqref{eq:x2} and dividing by $\langle T\rangle^2$ yields the equation for timing noise:
\begin{equation}\label{eq:cv2lin}
    CV^2_T\approx\frac{\langle b^2_2\rangle}{k_2\langle b_2\rangle^2\langle T\rangle}+\frac{X_1}{k_1\langle b_1\rangle\langle T\rangle^2}\left(\frac{\langle b^2_1\rangle}{k_1\langle b_1\rangle^2}-\frac{\langle b^2_2\rangle}{ k_2\langle b_2\rangle^2}\right).
\end{equation}
The expression \eqref{eq:cv2lin} is a linear function of the threshold $X_1$.~Therefore, the optimal value of $X_1$ that minimizes timing noise depends on the sign $\frac{\langle b^2_1\rangle}{k_1\langle b_1\rangle^2}-\frac{\langle b^2_2\rangle}{ k_2\langle b_2\rangle^2}$.~These conditions yield two possible scenarios of noise optimization:
\begin{itemize}
    \item \textbf{Using only the first gene minimizes timing noise $\left(\frac{\langle b^2_1\rangle}{k_1\langle b_1\rangle^2}<\frac{\langle b^2_2\rangle}{ k_2\langle b_2\rangle^2}\right)$:} In this scenario, the optimal thresholds are $(X_1^*=k_1\langle b_1\rangle\langle T\rangle, X_2^*=0)$, i.~e., using only the first gene.~Once the intrinsic noise of the first gene is lower than that of the second, adding a second protein accumulation process with a noisier gene will add more timing noise to the system.
    \item \textbf{Using only the second gene minimizes timing noise $\left(\frac{\langle b^2_1\rangle}{k_1\langle b_1\rangle^2}>\frac{\langle b^2_2\rangle}{ k_2\langle b_2\rangle^2}\right)$:} In this scenario, the optimal thresholds are $(X_1^*=0, X_2^*=k_2\langle b_2\rangle\langle T\rangle)$, i.~e., using only the second gene.~This means that the first gene is too noisy and using only the second gene, which has lower intrinsic noise, will minimize timing noise.
\end{itemize}

In the absence of protein dilution, the optimal solution consists of using the least noisy gene.~This can be concluded since the protein noise is given by \eqref{eq3c} and, therefore, the ratio $\frac{\langle b^2_2\rangle}{k_2\langle b_2\rangle^2}/\frac{\langle b^2_1\rangle}{ k_1\langle b_1\rangle^2}$ is related to the ratio of intrinsic noises of each gene $CV_2/CV_1$ through the following expression:
\begin{equation}
    \frac{\langle b^2_2\rangle}{k_2\langle b_2\rangle^2}/\frac{\langle b^2_1\rangle}{ k_1\langle b_1\rangle^2}=\left(CV_2/CV_1\right)^2
\end{equation}
Therefore, in the first scenario, we have ${CV_1/CV_2<1}$ and, similarly, in the second scenario ${CV_1/CV_2>1}$.~In the general case ($N>2$), the optimization process will be similar to the case of two genes, optimizing each time a new gene is added to the cascade.~Therefore, in the limit $\gamma \to 0$, the optimal strategy for $N > 1$ is using only the gene with the smallest intrinsic noise.

In the case of identical genes, using any number of genes will result in the same FPT noise as the case for one gene.~Using the system \eqref{eq:momsol2}, the FPT noise results in the expression:
\begin{equation}\label{eq:limit}
\left(CV^2_T\right)_{\text{identical}}=\frac{\langle b_1^2\rangle}{\langle b_1\rangle^2}\frac{1}{k_1\langle T\rangle}.
\end{equation}
Notice that when comparing expressions \eqref{eq:cv2lin} and \eqref{eq:limit}, the noise of identical genes defines a lower limit for the FPT noise given $\langle T\rangle$.

\bibliography{apssamp}% Produces the bibliography via BibTeX.

@article{matthews2022cell,
  title={Cell cycle control in cancer},
  author={Matthews, Helen K and Bertoli, Cosetta and de Bruin, Robertus AM},
  journal={Nature Reviews Molecular Cell Biology},
  volume={23},
  number={1},
  pages={74--88},
  year={2022},
  publisher={Nature Publishing Group UK London}
}

@article{pavlou2025single,
  title={Single-cell data reveal heterogeneity of investment in ribosomes across a bacterial population},
  author={Pavlou, Antrea and Cinquemani, Eugenio and Pinel, Corinne and Giordano, Nils and Mathilde, Van Melle-Gateau and Mihalcescu, Irina and Geiselmann, Johannes and de Jong, Hidde},
  journal={Nature Communications},
  volume={16},
  number={1},
  pages={285},
  year={2025},
  publisher={Nature Publishing Group UK London}
}

@article{kalir2001ordering,
  title={Ordering genes in a flagella pathway by analysis of expression kinetics from living bacteria},
  author={McClure, J and Pabbaraju, K and Southward, C and Ronen, M and Leibler, S and Surette, MG},
  journal={Science},
  volume={292},
  number={5524},
  pages={2080--2083},
  year={2001},
  publisher={American Association for the Advancement of Science}
}

@article{co2017stochastic,
  title={Stochastic timing in gene expression for simple regulatory strategies},
  author={Co, Alma Dal and Lagomarsino, Marco Cosentino and Caselle, Michele and Osella, Matteo},
  journal={Nucleic Acids Research},
  volume={45},
  number={3},
  pages={1069--1078},
  year={2017},
  publisher={Oxford University Press}
}

@article{nachman2007dissecting,
  title={Dissecting timing variability in yeast meiosis},
  author={Nachman, Iftach and Regev, Aviv and Ramanathan, Sharad},
  journal={Cell},
  volume={131},
  number={3},
  pages={544--556},
  year={2007},
  publisher={Elsevier}
}

@article{rijal2022exact,
  title={Exact distribution of threshold crossing times for protein concentrations: Implication for biological timekeeping},
  author={Rijal, Krishna and Prasad, Ashok and Singh, Abhyudai and Das, Dibyendu},
  journal={Physical Review Letters},
  volume={128},
  number={4},
  pages={048101},
  year={2022},
  publisher={APS}
}

@article{Sukys_Grima_2025, title={Cell-cycle dependence of bursty gene expression: Insights from fitting mechanistic models to single-cell RNA-seq data}, volume={53}, DOI={10.1093/nar/gkaf295}, number={7}, journal={Nucleic Acids Research}, author={Sukys, Augustinas and Grima, Ramon}, year={2025}, month={Apr}}

@article{Weiße_Oyarzún_Danos_Swain_2015, title={Mechanistic links between cellular trade-offs, gene expression, and growth}, volume={112}, DOI={10.1073/pnas.1416533112}, number={9}, journal={Proceedings of the National Academy of Sciences}, author={Weiße, Andrea Y. and Oyarzún, Diego A. and Danos, Vincent and Swain, Peter S.}, year={2015}, month={Feb}}

@article{Klumpp_Zhang_Hwa_2009, title={Growth rate-dependent global effects on gene expression in bacteria}, volume={139}, DOI={10.1016/j.cell.2009.12.001}, number={7}, journal={Cell}, author={Klumpp, Stefan and Zhang, Zhongge and Hwa, Terence}, year={2009}, month={Dec}, pages={1366–1375}}

@article{Keren_vanDijk_Weingarten-Gabbay_Davidi_Jona_Weinberger_Milo_Segal_2015, title={Noise in gene expression is coupled to growth rate}, volume={25}, DOI={10.1101/gr.191635.115}, number={12}, journal={Genome Research}, author={Keren, Leeat and van Dijk, David and Weingarten-Gabbay, Shira and Davidi, Dan and Jona, Ghil and Weinberger, Adina and Milo, Ron and Segal, Eran}, year={2015}, month={Sep}, pages={1893–1902}}

@article{singh2010stochastic,
  title={Stochastic hybrid systems for studying biochemical processes},
  author={Singh, Abhyudai and Hespanha, Joao P},
  journal={Philosophical Transactions of the Royal Society A: Mathematical, Physical and Engineering Sciences},
  volume={368},
  number={1930},
  pages={4995--5011},
  year={2010},
  publisher={The Royal Society Publishing}
}

@article{biswas2016redundancy,
  title={Redundancy in information transmission in a two-step cascade},
  author={Biswas, Ayan and Banik, Suman K},
  journal={Physical Review E},
  volume={93},
  number={5},
  pages={052422},
  year={2016},
  publisher={APS}
}

@article{pedraza2007random,
  title={Random timing in signaling cascades},
  author={Pedraza, Juan M and Paulsson, Johan},
  journal={Molecular systems biology},
  volume={3},
  number={1},
  pages={81},
  year={2007},
  publisher={John Wiley \& Sons, Ltd Chichester, UK}
}

@article{mondal2024molecular,
  title={Molecular mechanisms of precise timing in cell lysis},
  author={Mondal, Anupam and Teimouri, Hamid and Kolomeisky, Anatoly B},
  journal={Biophysical Journal},
  volume={123},
  number={18},
  pages={3090--3099},
  year={2024},
  publisher={Elsevier}
}

@article{kannoly2020optimum,
  title={Optimum threshold minimizes noise in timing of intracellular events},
  author={Kannoly, Sherin and Gao, Tianhui and Dey, Supravat and Wang, Nang and Singh, Abhyudai and Dennehy, John J},
  journal={iScience},
  volume={23},
  number={6},
  year={2020},
  publisher={Elsevier}
}

@inproceedings{dey2021feedforward,
  title={Feedforward genetic circuits regulate the precision of event timing},
  author={Dey, Supravat and Kannoly, Sherin and Bokes, Pavol and Dennehy, John J and Singh, Abhyudai},
  booktitle={2021 European Control Conference (ECC)},
  pages={2127--2132},
  year={2021},
  organization={IEEE}
}

@article{kannoly2022optimal,
  title={An optimal lysis time maximizes bacteriophage fitness in quasi-continuous culture},
  author={Kannoly, Sherin and Singh, Abhyudai and Dennehy, John J},
  journal={mBio},
  volume={13},
  number={3},
  pages={e03593--21},
  year={2022},
  publisher={Am Soc Microbiol}
}

@inproceedings{nieto2022threshold,
  title={Threshold-crossing time statistics for size-dependent gene expression in growing cells},
  author={Nieto, C{\'e}sar and Ghusinga, Khem Raj and Vargas-Garc{\'\i}a, C{\'e}sar and Singh, Abhyudai},
  booktitle={2022 American Control Conference (ACC)},
  pages={1341--1346},
  year={2022},
  organization={IEEE}
}

@article{Zhang_Zabaikina_Nieto_Vahdat_Bokes_Singh_2025, title={Stochastic gene expression in proliferating cells: Differing noise intensity in single-cell and Population Perspectives}, volume={21}, DOI={10.1371/journal.pcbi.1013014}, number={6}, journal={PLoS Computational Biology}, author={Zhang, Zhanhao and Zabaikina, Iryna and Nieto, Cesar and Vahdat, Zahra and Bokes, Pavol and Singh, Abhyudai}, year={2025}, month={Jun}}

@inproceedings{rezaee2023optimizing,
  title={Optimizing precision in cellular clocks through self-regulated accumulation of molecules},
  author={Rezaee, Sayeh and Nieto, C{\'e}sar and Singh, Abhyudai},
  booktitle={2023 27th International Conference on System Theory, Control and Computing (ICSTCC)},
  pages={505--510},
  year={2023},
  organization={IEEE}
}

@article{Garcia_Phillips_2011, title={Quantitative dissection of the simple repression input–output function}, volume={108}, DOI={10.1073/pnas.1015616108}, number={29}, journal={Proceedings of the National Academy of Sciences}, author={Garcia, Hernan G. and Phillips, Rob}, year={2011}, month={Jul}, pages={12173–12178}}

@article{Bintu_Buchler_Garcia_Gerland_Hwa_Kondev_Kuhlman_Phillips_2005, title={Transcriptional regulation by the numbers: Applications}, volume={15}, DOI={10.1016/j.gde.2005.02.006}, number={2}, journal={Current Opinion in Genetics \&; Development}, author={Bintu, Lacramioara and Buchler, Nicolas E and Garcia, Hernan G and Gerland, Ulrich and Hwa, Terence and Kondev, Jané and Kuhlman, Thomas and Phillips, Rob}, year={2005}, month={Apr}, pages={125–135}}

@article{Rao_Waxman_Lin_Song_2025, title={Exact first-passage time distributions from time-dependent solutions of the chemical master equation. I. Nonlinear Networks with bimolecular reactions and poisson-product initial conditions}, volume={162}, DOI={10.1063/5.0253020}, number={22}, journal={The Journal of Chemical Physics}, author={Rao, Changqian and Waxman, David and Lin, Wei and Song, Zhuoyi}, year={2025}, month={Jun}}

@article{Biswas_Dey_Singh_2024, title={Sequestration of gene products by decoys enhances precision in the timing of intracellular events}, volume={14}, DOI={10.1038/s41598-024-75505-y}, number={1}, journal={Scientific Reports}, author={Biswas, Kuheli and Dey, Supravat and Singh, Abhyudai}, year={2024}, month={Nov}}

@article{Ali_Prasad_Das_2025, title={Exact distributions of threshold crossing times of proteins under post-transcriptional regulation by small {RNA}S}, volume={111}, DOI={10.1103/physreve.111.014405}, number={1}, journal={Physical Review E}, author={Ali, Syed Yunus and Prasad, Ashok and Das, Dibyendu}, year={2025}, month={Jan}}

@article{ghusinga2017first,
  title={First-passage time approach to controlling noise in the timing of intracellular events},
  author={Ghusinga, Khem Raj and Dennehy, John J and Singh, Abhyudai},
  journal={Proceedings of the National Academy of Sciences},
  volume={114},
  number={4},
  pages={693--698},
  year={2017},
  publisher={National Acad Sciences}
}

@article{amir2007noise,
  title={Noise in timing and precision of gene activities in a genetic cascade},
  author={Amir, Amnon and Kobiler, Oren and Rokney, Assaf and Oppenheim, Amos B and Stavans, Joel},
  journal={Molecular Systems Biology},
  volume={3},
  number={1},
  pages={71},
  year={2007},
  publisher={John Wiley \& Sons, Ltd Chichester, UK}
}

@article{eldar2010functional,
  title={Functional roles for noise in genetic circuits},
  author={Eldar, Avigdor and Elowitz, Michael B},
  journal={Nature},
  volume={467},
  number={7312},
  pages={167--173},
  year={2010},
  publisher={Nature Publishing Group UK London}
}

@article{ali2022controlling,
  title={Controlling gene expression timing through gene regulatory architecture},
  author={Ali, Md Zulfikar and Brewster, Robert C},
  journal={PLoS Computational Biology},
  volume={18},
  number={1},
  pages={e1009745},
  year={2022},
  publisher={Public Library of Science San Francisco, CA USA}
}

@article{thattai2002attenuation,
  title={Attenuation of noise in ultrasensitive signaling cascades},
  author={Thattai, Mukund and van Oudenaarden, Alexander},
  journal={Biophysical Journal},
  volume={82},
  number={6},
  pages={2943--2950},
  year={2002},
  publisher={Elsevier}
}

@article{Ozbudak_Thattai_Kurtser_Grossman_vanOudenaarden_2002, title={Regulation of noise in the expression of a single gene}, volume={31}, DOI={10.1038/ng869}, number={1}, journal={Nature Genetics}, author={Ozbudak, Ertugrul M. and Thattai, Mukund and Kurtser, Iren and Grossman, Alan D. and van Oudenaarden, Alexander}, year={2002}, month={Apr}, pages={69–73}}

@article{Pedraza_Paulsson_2008, title={Effects of molecular memory and bursting on fluctuations in gene expression}, volume={319}, DOI={10.1126/science.1144331}, number={5861}, journal={Science}, author={Pedraza, Juan M. and Paulsson, Johan}, year={2008}, month={Jan}, pages={339–343}}

@article{Cao_Qiu_Zhou_Zhang_2019, title={Control Strategies for the timing of intracellular events}, volume={100}, DOI={10.1103/physreve.100.062401}, number={6}, journal={Physical Review E}, author={Cao, Mengfang and Qiu, Baohua and Zhou, Tianshou and Zhang, Jiajun}, year={2019}, month={Dec}}

@article{Thattai_vanOudenaarden_2001, title={Intrinsic noise in gene regulatory networks}, volume={98}, DOI={10.1073/pnas.151588598}, number={15}, journal={Proceedings of the National Academy of Sciences}, author={Thattai, Mukund and van Oudenaarden, Alexander}, year={2001}, month={Jul}, pages={8614–8619}}

@article{ng2024role,
  title={Role of Mitogen-Activated Protein (MAP) Kinase Pathways in Metabolic Diseases},
  author={Ng, Gavin Yong Quan and Loh, Zachary Wai-Loon and Fann, David Y and Mallilankaraman, Karthik and Arumugam, Thiruma V and Hande, M Prakash},
  journal={Genome Integrity},
  volume={15},
  year={2024},
  publisher={ScienceOpen}
}

@article{english2021designing,
  title={Designing biological circuits: synthetic biology within the operon model and beyond},
  author={English, Max A and Gayet, Rapha{\"e}l V and Collins, James J},
  journal={Annual Review of Biochemistry},
  volume={90},
  number={1},
  pages={221--244},
  year={2021},
  publisher={Annual Reviews}
}

@article{singh2014stochastic,
  title={Stochastic holin expression can account for lysis time variation in the bacteriophage $\lambda$},
  author={Singh, Abhyudai and Dennehy, John J},
  journal={Journal of the Royal Society Interface},
  volume={11},
  number={95},
  pages={20140140},
  year={2014},
  publisher={The Royal Society}
}

@article{avruch2007map,
  title={MAP kinase pathways: the first twenty years},
  author={Avruch, Joseph},
  journal={Biochimica et Biophysica Acta (BBA)-Molecular Cell Research},
  volume={1773},
  number={8},
  pages={1150--1160},
  year={2007},
  publisher={Elsevier}
}

@article{Cai_Friedman_Xie_2006, title={Stochastic protein expression in individual cells at the single molecule level}, volume={440}, DOI={10.1038/nature04599}, number={7082}, journal={Nature}, author={Cai, Long and Friedman, Nir and Xie, X. Sunney}, year={2006}, month={Mar}, pages={358–362}}

@article{Golding_Paulsson_Zawilski_Cox_2005, title={Real-time kinetics of gene activity in individual bacteria}, volume={123}, DOI={10.1016/j.cell.2005.09.031}, number={6}, journal={Cell}, author={Golding, Ido and Paulsson, Johan and Zawilski, Scott M. and Cox, Edward C.}, year={2005}, month={Dec}, pages={1025–1036}}

@article{grima_fpt,
  title={A stochastic vs deterministic perspective on the timing of cellular events},
  author={Ham, Lucy and Coomer, Megan A and {\"O}cal, Kaan and Grima, Ramon and Stumpf, Michael PH},
  journal={Nature Communications},
  volume={15},
  number={1},
  pages={5286},
  year={2024},
  publisher={Nature Publishing Group UK London}
}

@article{Elowitz_Levine_Siggia_Swain_2002, title={Stochastic gene expression in a single cell}, volume={297}, DOI={10.1126/science.1070919}, number={5584}, journal={Science}, author={Elowitz, Michael B. and Levine, Arnold J. and Siggia, Eric D. and Swain, Peter S.}, year={2002}, month={Aug}, pages={1183–1186}}

@article{friedman2006linking,
  author  = {Friedman, Nir and Cai, Long and Xie, X. Sunney},
  title   = {Linking stochastic dynamics to population distribution: an analytical framework of gene expression},
  journal = {Physical Review Letters},
  year    = {2006},
  volume  = {97},
  number  = {16},
  pages   = {168302},
  doi     = {10.1103/PhysRevLett.97.168302},
  url     = {https://link.aps.org/doi/10.1103/PhysRevLett.97.168302}
}

@article{shahrezaei2008analytical,
  author  = {Shahrezaei, Vahid and Swain, Peter S.},
  title   = {Analytical distributions for stochastic gene expression},
  journal = {Proceedings of the National Academy of Sciences},
  year    = {2008},
  volume  = {105},
  number  = {45},
  pages   = {17256--17261},
  doi     = {10.1073/pnas.0803850105},
  url     = {https://www.pnas.org/doi/10.1073/pnas.0803850105}
}

@article{Yu_Xiao_Ren_Lao_Xie_2006, title={Probing gene expression in live cells, one protein molecule at a time}, volume={311}, DOI={10.1126/science.1119623}, number={5767}, journal={Science}, author={Yu, Ji and Xiao, Jie and Ren, Xiaojia and Lao, Kaiqin and Xie, X. Sunney}, year={2006}, month={Mar}, pages={1600–1603}}

@article{Chong_Chen_Ge_Xie_2014, title={Mechanism of transcriptional bursting in bacteria}, volume={158}, DOI={10.1016/j.cell.2014.05.038}, number={2}, journal={Cell}, author={Chong, Shasha and Chen, Chongyi and Ge, Hao and Xie, X. Sunney}, year={2014}, month={Jul}, pages={314–326}}

@article{Fukaya_Lim_Levine_2016, title={Enhancer control of transcriptional bursting}, volume={166}, DOI={10.1016/j.cell.2016.05.025}, number={2}, journal={Cell}, author={Fukaya, Takashi and Lim, Bomyi and Levine, Michael}, year={2016}, month={Jul}, pages={358–368}}

@article{Rodriguez_Larson_2020, title={Transcription in living cells: Molecular mechanisms of bursting}, volume={89}, DOI={10.1146/annurev-biochem-011520-105250}, number={1}, journal={Annual Review of Biochemistry}, author={Rodriguez, Joseph and Larson, Daniel R.}, year={2020}, month={Jun}, pages={189–212}}

@article{gillespie2009deterministic,
  title={Deterministic limit of stochastic chemical kinetics},
  author={Gillespie, Daniel T},
  journal={The Journal of Physical Chemistry B},
  volume={113},
  number={6},
  pages={1640--1644},
  year={2009},
  publisher={ACS Publications}
}

@article{gupta2018temporal,
  title={Temporal precision of regulated gene expression},
  author={Gupta, Shivam and Varennes, Julien and Korswagen, Hendrik C and Mugler, Andrew},
  journal={PLoS computational biology},
  volume={14},
  number={6},
  pages={e1006201},
  year={2018},
  publisher={Public Library of Science San Francisco, CA USA}
}

@article{gupta2020temporal,
  title={Temporal precision of molecular events with regulation and feedback},
  author={Gupta, Shivam and Fancher, Sean and Korswagen, Hendrik C and Mugler, Andrew},
  journal={Physical Review E},
  volume={101},
  number={6},
  pages={062420},
  year={2020},
  publisher={APS}
}

@inproceedings{rezaee2024controlling,
  title={Controlling biomolecular timekeeping via regulated gene product degradation},
  author={Rezaee, Sayeh and Nieto, Cesar and Singh, Abhyudai},
  booktitle={2024 IEEE 63rd Conference on Decision and Control (CDC)},
  pages={7772--7777},
  year={2024},
  organization={IEEE}
}

@inproceedings{nieto2022feedback,
  title={Feedback strategies for threshold crossing of protein levels at a prescribed time},
  author={Nieto, C{\'e}sar and Ghusinga, Khem Raj and Singh, Abhyudai},
  booktitle={2022 30th Mediterranean Conference on Control and Automation (MED)},
  pages={170--175},
  year={2022},
  organization={IEEE}
}

@article{Gupta_Johnson_Cruz_Costa_Guest_Li_Hart_Nguyen_Stadlmeier_Bratton_2024, title={Global protein turnover quantification in \textit{Escherichia coli} reveals cytoplasmic recycling under nitrogen limitation}, volume={15}, DOI={10.1038/s41467-024-49920-8}, number={1}, journal={Nature Communications}, author={Gupta, Meera and Johnson, Alex N. and Cruz, Edward R. and Costa, Eli J. and Guest, Randi L. and Li, Sophia Hsin-Jung and Hart, Elizabeth M. and Nguyen, Thao and Stadlmeier, Michael and Bratton, Benjamin P. and et al.}, year={2024}, month={Jul}}

\end{document}